\documentclass[%
 reprint,
 amsmath,amssymb,
 aps,
]{revtex4-2}

\usepackage{graphicx}% Include figure files
\usepackage{dcolumn}% Align table columns on decimal point
\usepackage{bm}% bold math
\begin{document}

%\preprint{APS/123-QED}

\title{Chisel-edged screens to reduce loss in highly overmoded \\ THz iris lines with finite screen thickness}% Force line breaks with \\

\author{Adham Naji$^{1,2}$, Karl Bane$^{2}$, Andrei Trebushinin$^{3}$, and Paul Warr$^{4}$}

  \affiliation{$^1$\mbox{Santa Clara University, Santa Clara, CA, 95053} \\ $^2$\mbox{SLAC National Linear Accelerator Laboratory, Stanford University, Menlo Park, CA 94025} \\ $^3$\mbox{European~XFEL~GmbH,~Holzkoppel~4,~22869~Schenefeld,~Germany}  \\ $^4$\mbox{University of Bristol, Bristol, BS8 1TH, UK}}
  
\thanks{anaji@scu.edu}
  
\date{\today} % It is always \today, today,

\begin{abstract}
In this note we report observations on the growth trends of diffraction power loss and ohmic power loss, as functions of screen thickness, in highly overmoded THz iris-line waveguides that are constructed from thin screens. Recent theoretical developments have given a detailed field description, including eigen and transient analyses, that characterizes such waveguides under paraxial dipole-mode excitation. Informed by these analyses, we can better estimate -- and minimize -- power loss effects due to finite screen thickness in practical realizations. A geometric variation is proposed whereby we limit the growth trend in ohmic loss, even when using relatively thicker screens in practice. 
\end{abstract}

\maketitle

%\tableofcontents

%=========================================================
%=========================================================
%=========================================================
%=========================================================

% =================================================================================
\section{Introduction and Context}

\begin{figure}
  \begin{center}
  \includegraphics[width=3.1in]{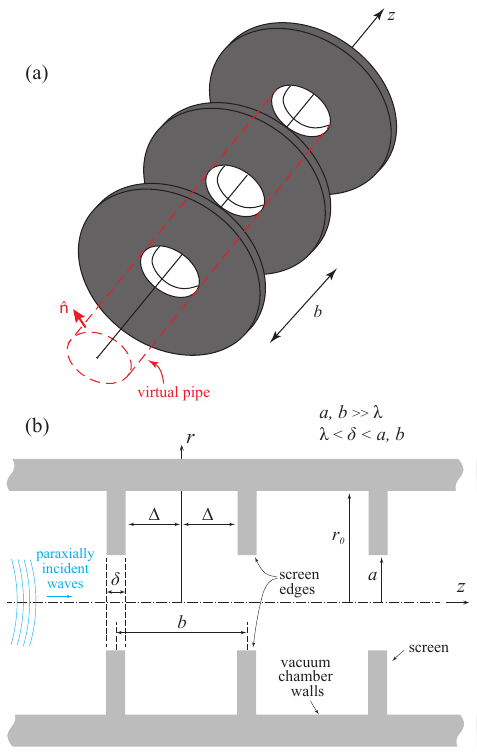}
 % \vspace{-15pt}
  \caption{A sketch of the iris-line waveguide in its regular configuration. (a) The perspective geometry without the vacuum chamber. (b) The cross-section, showing the chamber walls and the screen edges before chiseling. }\label{Figure1}
  \end{center}
\end{figure}

The use of highly overmoded iris-line waveguides for the efficient transportation of THz radiation over long distances (hundreds of meters) has been recently investigated \cite{NajiTHz1, NajiTHz2, DESY}. One of the main motivations behind such an investigation is the need for a geometrically simple and relatively inexpensive structure that can offer low-loss transport for paraxially incident THz waves, with uniform linear (dipole) polarization and approximately Gaussian transverse intensity profile, at the SLAC National Accelerator Laboratory, Stanford University \cite{DESY, NajiTHz1,NajiTHz2, Zhang}. Two theoretical studies, a steady-state eigen-mode field analysis and a transient-regime field analysis \cite{NajiTHz1, NajiTHz2}, with their associated computer codes, were developed and have highlighted the main features of wave propagation in the iris-line structure, with experimental demonstrations currently underway at SLAC. Figure~\ref{Figure1} shows the basic structure under consideration as a candidate THz transportation method for SLAC's Linac Coherent Light Source (LCLS-II) facility. A new THz wiggler device, to be installed downstream of the LCLS-II soft x-ray undulators, will generate radiation in the 3--15~THz range, for use in future pump-probe experiments \cite{Zhang}. In such applications, short pulses (e.g.~few picoseconds long) that carry the THz radiation require transportation over a distance of 150--300~m. Adoption of the iris-line structure for similar applications at the European X-Ray Free Electron Laser (European XFEL) facility in Germany is also under consideration \cite{DESY3,DESY4}.

Periodic structures like the iris line have been well studied in classical literature on microwave/RF waveguides (e.g.~see \cite{Borgnis}). What makes the THz iris line under current investigation different, however, is its highly oversized dimensions relative to the THz wavelength ($\lambda {<} 100~\mu$m) and the fact that its ``thin" screens ($\delta{<}a,b$ in Figure~\ref{Figure1}) are actually of non-negligible thickness relative to the wavelength ($\delta >\lambda$). In contrast to traditional microwave corrugated waveguides (e.g.~\cite{Mahmoud,Collin}) or slow-wave RF accelerator structures (e.g.~\cite{slaterbook,WanglerBook}), where the structure's period and transverse dimensions are typically sub-wavelength in size, the present structure has a size of the order of hundreds or thousands of wavelengths and can host hundreds of modes (highly overmoded). Diffraction loss, which here has the attractive property of scaling inversely with the cubic power of the iris radius (${\sim}1/a^3$), dominates the transmission loss in such overmoded structures when thin screens are used \cite{NajiTHz1,NajiTHz2,Vainstein1,Vainstein2}, whereas ohmic loss is the main contributor to power loss in traditional sub-wavelength corrugated structures \cite{Mahmoud}. 

Traditional models typically analyze such overmoded structures assuming zero thickness ($\delta\ll\lambda$); an assumption that is helpful to reduce the stringent computational burden when modeling such geometries, but one that is no longer warranted for accurate modeling when practical screen thickness is larger than $\lambda$ itself. The field analyses in \cite{NajiTHz1} and \cite{NajiTHz2} were developed to include the effects of finite screen thickness into account. We have shown that diffraction loss is relatively decreased when a thin-but-finite screen thickness is used (i.e. having $\delta < a,b$ and $\delta >\lambda$ simultaneously), but remains dominant compared to ohmic loss from the screen edges \cite{NajiTHz1,NajiTHz2}. On the other hand, ohmic loss, rising chiefly from longitudinal and azimuthal surface currents on the rims of the irises due to tangential magnetic fields, tends to increase with thicker screens.  

In this paper, we highlight a design technique that uses single-sided chiseled screen edges as a means of keeping ohmic loss as low as possible, while being able to utilize relatively thicker screen, which tend to be more convenient in practice. The technique is informed by the analytical modeling developed in \cite{NajiTHz1} and \cite{NajiTHz2}, which helps our understanding of the wave phenomena taking place in the iris-line structure. 

The rest of the paper is organized as follows. Section~\ref{sec:FieldEquations} briefly reviews the field equations that govern the physics of the structure. The implication of using thicker screens on ohmic and diffraction losses in the iris line are discussed in Section~\ref{sec:LossTrends}. The proposed technique of using single-sided chiseled edges is then described in Section~\ref{sec:Chisel} and the paper is concluded in Section~\ref{sec:Conclusions}.

% =================================================================================
% =================================================================================
\section{Field Description} \label{sec:FieldEquations}
In the steady-state, the dipolar eigenmode propagating in the iris line with screens of nonzero thickness is dominated by a hybrid mode that can be computed using the perturbative (clustered) mode-matching technique developed in \cite{NajiTHz1}. This can be expressed by a spatial Fourier expansion as
\begin{align} 
E_{r}&=i\cos \phi \sum\limits_{n=-\infty}^{\infty} \frac{D_{n}\omega\mu}{Z_{0}rk^{2}_{tn}}\frac{J_{1}(k_{tn}r)}{J_{1}(k_{tn}a)} {+} \frac{C_{n}\beta_{n}}{k_{tn}} \frac{J'_{1}(k_{tn}r)}{J_{1}(k_{tn}a)} e^{i\beta_{n}z}\nonumber\\
E_{\phi}&=i\sin \phi \sum\limits_{n=-\infty}^{\infty}  {-}\frac{D_{n}\omega\mu}{Z_{0}k_{tn}}\frac{J'_{1}(k_{tn}r)}{J_{1}(k_{tn}a)} {-}  \frac{C_{n}\beta_{n}}{r k^{2}_{tn}} \frac{J_{1}(k_{tn}r)}{J_{1}(k_{tn}a)} e^{i\beta_{n}z}\nonumber\\
E_{z}&=\cos\phi \sum\limits_{n=-\infty}^{\infty}C_{n} \frac{J_{1}(k_{tn}r)}{J_{1}(k_{tn}a)}e^{i\beta_{n}z} \nonumber\\
H_{r}&=i\sin \phi \sum\limits_{n=-\infty}^{\infty}   \frac{D_{n}\beta_{n}}{Z_{0}k_{tn}}\frac{J'_{1}(k_{tn}r)}{J_{1}(k_{tn}a)} {+}  \frac{C_{n}\omega\epsilon}{r k^{2}_{tn}} \frac{J_{1}(k_{tn}r)}{J_{1}(k_{tn}a)} e^{i\beta_{n}z}\nonumber\\
H_{\phi}&=i\cos \phi \sum\limits_{n=-\infty}^{\infty} \frac{D_{n}\beta_{n}}{Z_{0}rk^{2}_{tn}}\frac{J_{1}(k_{tn}r)}{J_{1}(k_{tn}a)} {+}  \frac{C_{n}\omega\epsilon}{k_{tn}} \frac{J'_{1}(k_{tn}r)}{J_{1}(k_{tn}a)}e^{i\beta_{n}z} \nonumber\\
H_{z}&=\sin \phi \sum\limits_{n=-\infty}^{\infty}\frac{D_{n}}{Z_{0}} \frac{J_{1}(k_{tn}r)}{J_{1}(k_{tn}a)}e^{i\beta_{n}z}\nonumber
\end{align}
where, for the $n$th harmonic, $k_{tn}{=}\sqrt{k^{2}-\beta_{n}^{2}}$ is the transverse wavenumber, $\beta_{n}=k+2\pi n/b$ is the harmonic wavenumber, $k{=}\omega/c$ is the wavenumber in free-sapce, $Z_{0}, \epsilon, \mu$ are the impedance, permittivity and permeability of free-space, respectively, $C_{n}, D_{n}$ are arbitrary amplitude coefficients (see \cite{NajiTHz1} for details on how to compute them), $J_{n}$ is the Bessel function of the first kind and $n$th order and $J'_{n}$ is its derivative with respect to its argument. It can be shown (see Appendix A in \cite{NajiTHz1}) that in the limit of infinitely thin screens ($\delta\rightarrow 0$), the transverse electric field profile reduces to the desirable form found in \cite{DESY} and \cite{Vainstein1}, which can be written as $E_{\perp}\sim J_{0}\left[\frac{2.4 r}{a}\left(1-\varepsilon-i\varepsilon\right)\right]$, where $\varepsilon\approx0.164\sqrt{\lambda b}/a$.
% ==== FIG =========
% ==== ==== =========
 The profile of this mode is akin to the $J_{0}(2.4r/a)$ profile, but with a small perturbation $\varepsilon$ that enters through its complex argument. The latter is a key feature of this system and represents the complex impedance equivalent to the structure's periodic nature, as seen by the hybrid mode. The uniform linear polarizarion across the iris is also a result of this boundary, which does not force the field to meet the wall in the same way a pure TE or TM mode would meet an ideal conductor. This field in the above equations represents the ``settled" mode in a very long iris line (like an ideal Bloch wave), after it has cleared any transient regime which may have formed at the line's entrance. The length of such a transient regime will not only depend on the structure properties, but also on the type of exciting wave incident on the input of the line (i.e.~how well-matched it is to the eigenmode). 
 
 A second description that can capture the transient regime can be obtained using mode-matching for the local field scattering that takes place down the line on a cell-by-cell basis (without exploiting the Bloch wave formalism) \cite{NajiTHz2}. This can be done using expansions in the TE and TM paraxial modes of the following form (given below for the sections with radius $a$; exchange $a\leftrightarrow r_{0}$ for fields in sections with radius $r_0$)
\begin{align}
   E_{ r_\text{TE}}&=\cos\phi\sum\limits_{n=1}^{\infty}\frac{A_{n}a}{\nu'_{1n} r} J_{1}(\nu'_{1n}\frac{ r}{a}) \ e^{-iz\frac{\nu'^{2}_{1n}}{2ka^{2}}+ikz}\nonumber\\
   E_{\phi_\text{TE}}&= -\sin\phi\sum\limits_{n=1}^{\infty}A_{n} J'_{1}(\nu'_{1n}\frac{ r}{a}) \ e^{-iz\frac{\nu'^{2}_{1n}}{2ka^{2}}+ikz}\nonumber\\
    H_{ r_\text{TE}}&=-\sin\phi\sum\limits_{n=1}^{\infty}\frac{A_{n}}{Z_{0}}\left(\frac{\nu'^{2}_{1n}}{2k^{2}a^{2}}-1 \right)J'_{1}(\nu'_{1n}\frac{ r}{a}) \ e^{-iz\frac{\nu'^{2}_{1n}}{2ka^{2}}+ikz}\nonumber\\
   H_{\phi_\text{TE}}&=-\cos\phi\sum\limits_{n=1}^{\infty}\frac{A_{n}a}{Z_{0} r\nu'_{1n}}\left(\frac{\nu'^{2}_{1n}}{2k^{2}a^{2}}-1 \right)J_{1}(\nu'_{1n}\frac{ r}{a}) \ e^{-iz\frac{\nu'^{2}_{1n}}{2ka^{2}}+ikz}\nonumber
      \end{align}
   \begin{align}
   H_{z_\text{TE}}&=-\sin\phi\sum\limits_{n=1}^{\infty}\frac{iA_{n}\nu'_{1n}}{Z_{0}k a}J_{1}(\nu'_{1n}\frac{ r}{a}) \ e^{-iz\frac{\nu'^{2}_{1n}}{2ka^{2}}+ikz}\nonumber\\
      E_{ r_\text{TM}}&=-\cos\phi\sum\limits_{n=1}^{\infty}B_{n}J'_{1}\left( \nu_{1n}\frac{ r}{a}\right) \ e^{-iz\frac{\nu^{2}_{1n}}{2ka^{2}}+ikz}\nonumber\\
   E_{\phi_\text{TM}}&= \sin\phi\sum\limits_{n=1}^{\infty}\frac{B_{n}a}{\nu_{1n} r}J_{1}\left( \nu_{1n}\frac{ r}{a}\right) \ e^{-iz\frac{\nu^{2}_{1n}}{2ka^{2}}+ikz}\nonumber\\
   E_{z_\text{TM}}&=\cos\phi\sum\limits_{n=1}^{\infty}\frac{iB_{n}\nu_{1n}}{ak}J_{1}\left( \nu_{1n}\frac{ r}{a} \right) \ e^{-iz\frac{\nu^{2}_{1n}}{2ka^{2}}+ikz}\nonumber\\
   H_{ r_\text{TM}}&=-\sin\phi\sum\limits_{n=1}^{\infty}\frac{B_{n}}{Z_{0} r}\left(\frac{\nu_{1n}}{2ak^{2}}+\frac{a}{\nu_{1n}} \right)J_{1}(\nu_{1n}\frac{ r}{a}) \ e^{-iz\frac{\nu^{2}_{1n}}{2ka^{2}}+ikz}\nonumber\\
   H_{\phi_\text{TM}}&=-\cos\phi\sum\limits_{n=1}^{\infty}\frac{B_{n}}{Z_{0}}\left(1+\frac{\nu^{2}_{1n}}{2a^{2}k^{2}}\right)J'_{1}(\nu_{1n}\frac{ r}{a}) \ e^{-iz\frac{\nu^{2}_{1n}}{2ka^{2}}+ikz}\nonumber
\end{align}
where $\nu_{1n}$ and $\nu'_{1n}$ denote the $n$th zero of the function $J_{1}$ or $J'_{1}$, respectively. The analytical and numerical calculation of the coefficients $A_{n}$ and $B_{n}$, for all TE/TM modes and discontinuities within the structure's cell, is discussed and demonstrated for various excitation scenarios in reference \cite{NajiTHz2}.

% =================================================================================
% =================================================================================
\section{Growth Trends in Loss for Thicker Screens} \label{sec:LossTrends}
\begin{figure}
  \begin{center}
  \includegraphics[width=3.6in]{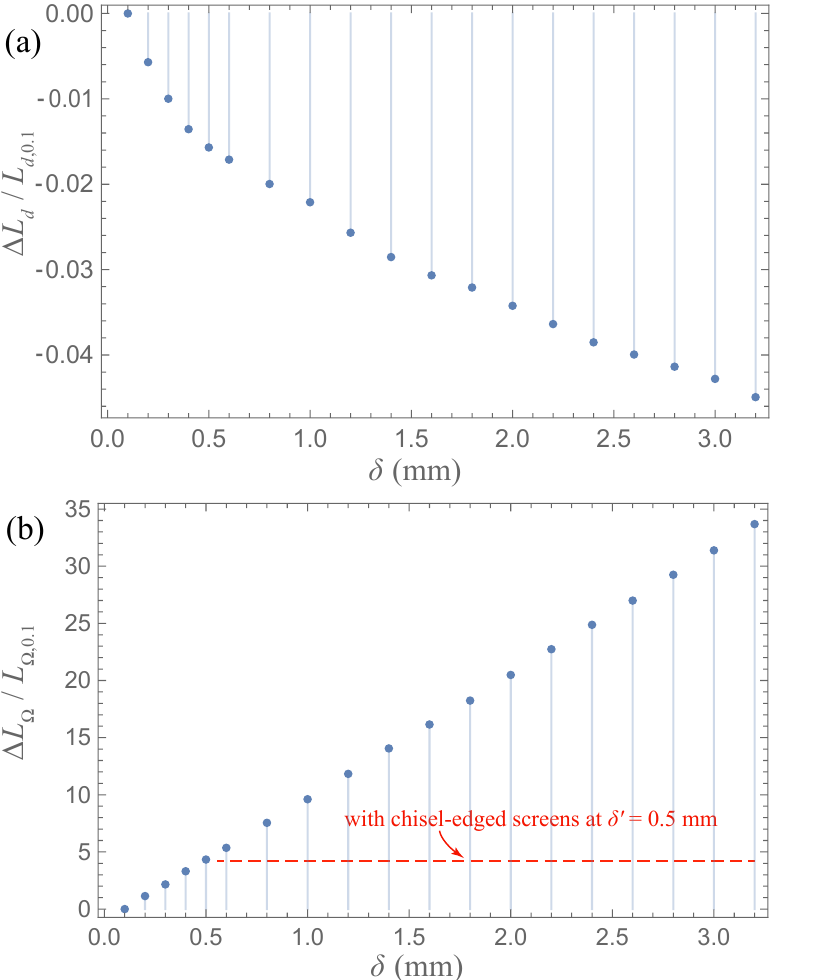}
 % \vspace{-15pt}
  \caption{Relative growth trends in diffraction and ohmic losses, as functions of thin screen thickness $\delta$. (a) Trend in diffraction loss relative to its value at a reference thickness, taken here at $\delta=0.1$~mm. (b) The corresponding relative growth in ohmic loss at screen edges (almost linear). Data shown for an iris line excited with a nearly matched $J_{0}(2.4r/a)$ dipole mode incidence at 3~THz and using copper screens with $a=5.5$~cm, $r_{0}=2a$ and $b=33.3$~cm. The red dashed line in (b) represents the predicted behavior of fixed ohmic loss if the screen edges were chispeled with a tip width $\delta'=0.5$, for instance, and $\delta>\delta'$. Note the expected effect of the chisel in fixing (capping) the growth of ohmic loss. }\label{Figure2} 
  \end{center}
\end{figure}

These field expansions for the iris line can be computed numerically to study the influence of an increasing screen thickness $\delta$ in practice. Within the limit of thin screens $\delta<b,a$ the diffraction loss remains dominant. For example, at 3~THz with nominal iris-line parameters $a=5.5$~cm, $r_{0}=2a$, $b=33.3$~cm and copper screens thinner than 2~mm, diffraction loss dominates the overall transmission power loss at approximately~15\% (ohmic loss is smaller than 1\%).  We generally observe, however, two basic trends when $\delta$ of a thin screen is increased: the diffraction loss ($\Delta L_{d}$) is relatively decreased, whereas the ohmic loss ($\Delta L_{\Omega}$) is relatively increased. Figure~\ref{Figure2} is a demonstration of these trends, as we increase $\delta$ for thin copper screens at 3~THz, with $a=5.5$~cm, $r_{0}=2a$ and $b=33.3$~cm. Note that the relative growth in the ohmic trend is faster than that for the diffraction trend for this highly overmoded system (a heavy-diffractive regime that can be roughly gauged by its high Frensnel number; $N_{f}=a^{2}/b \lambda=91$ in this example). Regardless of the specific parameters used, these relative trends are characteristic and expected to hold for any iris-line operating under a similarly overmoded state.

% ==== FIG =========
% ==== ==== =========

% ==== FIG =========
\begin{figure}
  \begin{center}
  \includegraphics[width=3.5in]{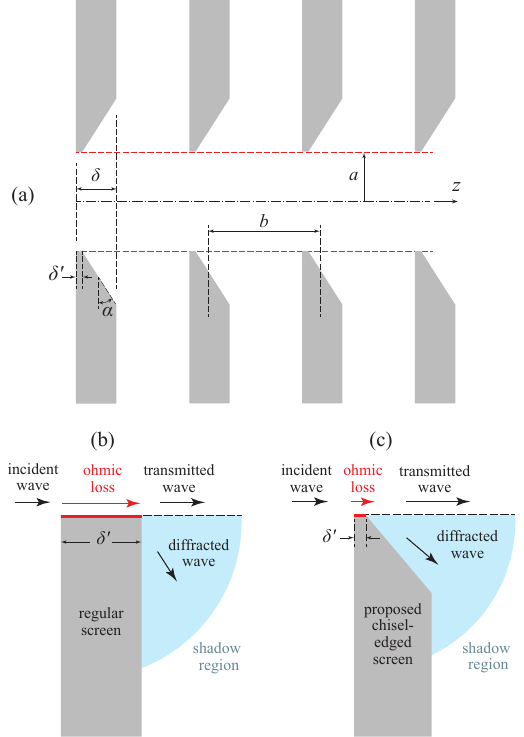}
 % \vspace{-15pt}
  \caption{Solution proposed to suppress the relative growth of ohmic loss at the screen edges. (a) The single-sided chisel-edged screens, with base thicknes $\delta$ and tip (edge) thickness $\delta'<\delta$. It is considered easier in practice to mechanically produce a chiseled cut of thin tip on a thicker screen, than to use a regular screen that is entirely as thin as the thin tip ($\delta'$). (b) and (c) show sketches of the concept of a chisel-edged screen, compared to a regular screen, and how it affects the edge area where ohmic losses are produced by the transmitted fields. The blue-shaded area represents the shadow region, where part of the signal's power is lost to diffraction between the screens.}\label{Figure3}
  \end{center}
\end{figure}
% ==== ==== =========

% =================================================================================
% =================================================================================
\section{Proposed Solution} \label{sec:Chisel}
Keeping ohmic loss as low as possible through the usage of extremely thin screens (as suggested by Figure~\ref{Figure3}b) is usually not convenient in practice due to limitations on mechanical production. To allow for relatively thicker screens to be used while keeping ohmic loss as low as possible, we propose cutting the iris edges obliquely near the top (like a chisel edge) and only from the side opposite to the source. These single-sided chisel-edged screens are drawn in Figure~\ref{Figure3}. This proposed technique is part of the configurations used in the experimental demonstrations currently underway at SLAC and European~XFEL for the THz iris line.

This technique allows the conductive surface responsible for the ohmic loss (namely, the rim of the iris) to shrink linearly in surface area, while still allowing the stem of the screen to be relatively wide. Joule's effect due to the longitudinal and azimuthal surface currents, $J_{s,z}$ and $J_{s,\phi}$, which are induced by the tangential magnetic field on the iris rim of width $\delta$, is thus almost linearly reduced with $\delta$. The chisel tip angle $\alpha$ is kept relatively sharp, so that the angle $\pi/2-\alpha$ remains larger than what the knife-edge diffraction effect would be for a corresponding plane wave; a first approximation of $\alpha\leq45^\circ$ is recommended (a detailed analysis including the effects of angles will be the subject of a future publication). Given that the screen edges are relatively thin and that any field power loss that happens between the screens (the shadow region in Figure~\ref{Figure3}) is already accounted for through the portion of power lost to diffraction, we expect a net reduction in ohmic loss at the edges at no considerable penalty in diffraction loss compared to the original edge design (see the dashed red line in Figure~\ref{Figure3}b). This fact also allows us to obtain a good modeling approximation by using the analytical equations developed for the efficient modeling of iris lines with regular screens (see \cite{NajiTHz1} and \cite{NajiTHz2}), but with thickness $\delta'$ instead of $\delta$. The side of the screen facing the incoming wave (the left side of the screens in the figures) is kept vertical to minimize any backward reflections toward the axial region of the waveguide, where they could interfere with the propagating wave.  

% =================================================================================
% =================================================================================

%=========================================================
%=========================================================

\section{\label{sec:Conclusions}Conclusions}
A simple practical solution has been proposed to minimze ohmic loss in highly overmoded THz iris lines when relatively thicker screens are used. The single-sided chisel-edged screens proposed in this paper offer the advantage of reducing ohmic loss, while maintaining the same diffraction loss performance and offering mechanical convenience. The analysis of the proposed variation is conveniently similar to the original analysis of regular (flat) screens, e.g.~as in \cite{NajiTHz1} and \cite{NajiTHz2}, further reducing the burden of optimization cycles. 

%=============================================================================
%=============================================================================
%=============================================================================
\begin{acknowledgments}
The authors thank Gennady Stupakov and Zhirong Huang, of SLAC National Accelerator Laboratory, Stanford University, for useful discussions related to the subject of this paper. Part of this work was supported
by the US Department of Energy (contract number DE-AC02-76SF00515).

\end{acknowledgments}

%=============================================================================
%=============================================================================
%=============================================================================
%\appendix

%===============================================================
%\section{\label{Appx1}Derivation of power loss, field polarization and field intensity properties based on Vainstein's boundary condition}

\hfill

%=============================================================================
%=============================================================================
%=============================================================================

\bibliography{myrefs}% Produces the bibliography via BibTeX.

\end{document}